\documentclass[preprints,article,accept,pdftex,moreauthors]{Definitions/mdpi} 
\firstpage{1} 
\pubvolume{1}
\issuenum{1}
\articlenumber{0}
\pubyear{2025}
\copyrightyear{2025}
\datereceived{ } 
\daterevised{ } 
\dateaccepted{ } 
\datepublished{ } 
\hreflink{https://doi.org/} 

\usepackage{amsmath}
\usepackage{amssymb}
\usepackage{slashed}
\usepackage{physics}

\Title{ Second-order pseudo-Hermitian spin-$1/2$ bosons}

\TitleCitation{Title}

\Author{Armando de la C. Rangel-Pantoja$^{1}$\orcidA{}, I. D\'iaz-Salda\~na $^{2}$\orcidB{} and Carlos A. Vaquera-Araujo $^{3,}$$^{1,}$$^{4,}$*\orcidC{}}

\AuthorNames{Armando de la C. Rangel-Pantoja, I. D\'iaz-Salda\~na and Carlos A. Vaquera-Araujo}

\AuthorCitation{Rangel-Pantoja, Armando de la C. .; D\'iaz-Salda\~na, I.; Vaquera-Araujo, C. A.}

\address{%
$^{1}$ \quad Departamento de F\'isica, División de Ciencias e Ingenierías, Campus Le\'on, Universidad de
  Guanajuato, Loma del Bosque 103, Lomas del Campestre, Le\'on, Guanajuato 37150, Mexico; \\
$^{2}$ \quad Unidad Acad\'emica de F\'isica, Universidad Aut\'onoma de Zacatecas, Calzada Solidaridad esquina con Paseo a la Bufa S/N, Zacatecas 98060, Zacatecas, Mexico; \\
$^{3}$ \quad Secretar\'ia de Ciencia, Humanidades, Tecnolog\'ia e Innovaci\'on, Insurgentes Sur 1582. Cr\'edito Constructor, Benito Ju\'arez 03940, Ciudad de M\'exico, Mexico;\\
$^{4}$ \quad Dual CP Institute of High Energy Physics, Colima 28045, Colima, Mexico.
}

\corres{Correspondence: vaquera@fisica.ugto.mx}

\abstract{The canonical quantization of a field theory for spin-$1/2$ massive bosons that satisfy the Klein-Gordon equation is presented. The breakdown of the usual spin-statistics connection is due to the redefinition of the dual field, rendering the theory pseudo-Hermitian. The normal-ordered Hamiltonian is bounded from below with real eigenvalues, and the theory is consistent with microcausality and invariant under parity, charge conjugation and time reversal.}


\begin{document}

\section{Introduction}

Spin-$1/2$ particles are successfully described at fundamental level by the Dirac equation, which is first-order in space-time derivatives and consistent with the Fermi-Dirac statistics observed in nature. In this formalism, any massive field that satisfies Dirac's equation simultaneously complies with the Klein-Gordon equation. Thus, one may ask the natural question: Can the dynamics of free massive fields be described by the Klein-Gordon equation with the correct treatment of the spin of the particles involved? The first second-order theory for spin-$1/2$ particles was formulated by Feynman in \cite{Feynman:1951gn}, following the work of Fock \cite{Fock:1937dy}. The V-A structure of weak interactions proposed by
Feynman and Gell-Mann was actually motivated by the existence of a second order equation
of motion for the fermionic degrees of freedom, which is easily handled by the method of path integrals, in contrast to the Dirac equation\cite{Feynman:1958ty}.  Thanks to this observation, a second-order formalism for spin-$1/2$ is especially useful in the world-line formulation of perturbative quantum field theory \cite{Schubert:2001he}. However, the quantization of the Lagrangian that generalizes the Klein-Gordon equation in this case  is not straightforward \cite{KP}. In this paper, we show that the canonical quantization of a Second-Order Pseudo-Hermitian theory (SOPHY) for massive bosons transforming under the $(\tfrac{1}{2},0)\oplus(0,\tfrac{1}{2})$ representation of the restricted Lorentz group (RLG) that obey the Klein-Gordon equation is indeed viable, and the key observation is to relax the requirement of Hermiticity in favor of pseudo-Hermiticiy. 

The absence of the usual spin-statistics connection in a pseudo-Hermitian quantum field theory was first studied in \cite{LeClair:2007iy}, where a second order formalism describing the dynamics of $N$-component complex symplectic fermions transforming as scalars in the $(0,0)$ representation of the RLG was discussed. Recently, there has been a renewed interest in pseudo-Hermitian theories with flipped statistics. In particular, in \cite{Ahluwalia:2022zrm} the quantization of spin-$1/2$ bosons satisfying the Dirac equation was studied, and in \cite{Ahluwalia:2023slc} a bosonic theory for Elko spinors with Wigner degeneracy was analyzed. In this work, we study the canonical quantization of the simplest Lorentz-invariant SOPHY for spin-$1/2$ bosons, with opposite statistics of the theory presented in \cite{Ferro-Hernandez:2023ymz}. The paper is structured as follows: In Section 2, the concept of a pseudo-Hermitian quantum field theory is presented. The canonical quantization of the theory is studied in Section 3. The symmetries of the theory are analyzed in Section 4, and finally, conclusions are drawn in Section 5. 

\section{Pseudo-hermitian theory}
A pseudo-Hermitian quantum field theory is described by a Lagrangian that satisfies 
\begin{equation}
\mathcal{L}^{\#}\equiv \eta^{-1}\mathcal{L}^\dagger\eta=\mathcal{L}.
\label{pseudo-Hermitian}
\end{equation}
for some operator $\eta$.
This departure of Hermiticity was studied in the context of quantum mechanics in \cite{Mostafazadeh:2001jk, Mostafazadeh:2008pw} generalizing the results of $PT$-symmetric systems introduced in \cite{Bender:1998ke}. Recently, there has been progress in the development of consistent pseudo-Hermitian quantum field theories \cite{Sablevice:2023odu}, in particular in pseudo-Hermitian quantum mechanics, there are two important results: 1) the energy spectrum of the theory is real and 2) the time evolution is unitary upon the definition of a suitable internal product of states based on the $\eta$ operator.

The Lagrangian proposed for the $\psi$ field transforming under the $(\tfrac{1}{2},0)\oplus(0,\tfrac{1}{2})$ of the RLG is given by
\begin{equation}\label{Lag1}
\begin{split}
\mathcal{L}= \partial^{\mu}\widehat{\psi}\partial_\mu\psi-m^2\widehat{\psi}\psi ,
\end{split}
\end{equation}
where $\widehat{\psi}$ is the modified dual field that renders the theory pseudo-Hermitian.
As pointed out in \cite{Ferro-Hernandez:2023ymz}, in contrast to Dirac theory, our second-order formalism describes a field with eight degrees of freedom and mass dimension one. The most general solution to the Klein-Gordon equation
for our spin-$1/2$ fields can be written in terms of two independent Dirac spinors $\psi_{1}$ and $\psi_2$ of the form \cite{CufaroPetroni:1985tu}
\begin{equation}
\psi=\frac{1}{\sqrt{2m}}(\psi_1+\gamma^5\psi_2),
\end{equation}
and the mode expansions of $\psi$ and $\widehat\psi$ are given by
\begin{align}
\psi(x)=&\int \frac{d^3\mathbf{p}}{(2\pi)^{3}2\sqrt{m\omega_{\mathbf{p}}}}\sum_s \bigg\{\left(u^{s}_{\mathbf{p}}a^{1 s}_{\mathbf{p}}+\gamma^5u^{s}_{\mathbf{p}}a^{2 s}_{\mathbf{p}}\right)e^{-ip\cdot x}+\left(v^{s}_{\mathbf{p}}b^{1 s\dagger}_{\mathbf{p}}+\gamma^5v^{s}_{\mathbf{p}}b^{2 s\dagger}_{\mathbf{p}}\right)e^{ip\cdot x}\bigg\},\nonumber\\
\widehat{\psi}(x)=&\int \frac{d^3\mathbf{p}}{(2\pi)^{3}2\sqrt{m\omega_{\mathbf{p}}}}\sum_s \bigg\{\left(\bar{u}^{s}_{\mathbf{p}}a^{1 s\dagger}_{\mathbf{p}}+\bar{u}^{s}_{\mathbf{p}}\gamma^5a^{2 s\dagger}_{\mathbf{p}}\right)e^{ip\cdot x}-\left(\bar{v}^{s}_{\mathbf{p}}b^{1 s}_{\mathbf{p}}+\bar{v}^{s}_{\mathbf{p}}\gamma^5b^{2 s}_{\mathbf{p}}\right)e^{-ip\cdot x}\bigg\},\label{field_expansion}
\end{align}
with $\omega_{\mathbf{p}}=+\sqrt{|\mathbf{p}|^2+m^2}$, $p^\mu=(\omega_{\mathbf{p}},\mathbf{p})$, and $s=\pm 1/2$. The positive and negative energy solutions of the Dirac free equation are
\begin{equation}\label{u_v}
u^{s}_{\mathbf{p}}=\begin{pmatrix}
\sqrt{p\cdot\sigma }\phi^s\\
\sqrt{p\cdot\overline{\sigma }}\phi^s
\end{pmatrix},\qquad 
v^{s}_{\mathbf{p}}=\begin{pmatrix}
\sqrt{p\cdot\sigma }\chi^s\\
-\sqrt{p\cdot\overline{\sigma }}\chi^s
\end{pmatrix},
\end{equation}
where $\sigma^\mu=(\mathbf{1},\boldsymbol{\sigma})$ and $\overline{\sigma}^\mu=(\mathbf{1},-\boldsymbol{\sigma})$. Here,  the reference Pauli spinors satisfy the relation $\chi^s=-i\sigma^2\phi^{s*}$, and are explicitly given by
\begin{equation}
\phi^{\frac{1}{2}}=\begin{pmatrix}
1\\
0
\end{pmatrix},\quad
\phi^{-\frac{1}{2}}=\begin{pmatrix}
0\\
1
\end{pmatrix},\quad
\chi^{\frac{1}{2}}=\begin{pmatrix}
0\\
1
\end{pmatrix},\quad
\chi^{-\frac{1}{2}}=\begin{pmatrix}
-1\\
0
\end{pmatrix}.
\end{equation} 
The spinors $u^{s}_{\mathbf{p}}$, $v^{s}_{\mathbf{p}}$ are normalized according to
\begin{equation}
\begin{array}{ccc}
\bar{u}^{r}_{\mathbf{p}}u^{s}_{\mathbf{p}}=2m\delta^{rs}, &\qquad \qquad&\bar{v}^{r}_{\mathbf{p}}v^{s}_{\mathbf{p}}=-2m\delta^{rs}, \\
\end{array}
\end{equation}
and are orthogonal to each other $\bar{u}^{r}_{\mathbf{p}}v^{s}_{\mathbf{p}}=\bar{v}^{r}_{\mathbf{p}}u^{s}_{\mathbf{p}}=0$.
Their completeness relations can be written as
\begin{equation}
\begin{array}{ccc}
\sum_{s} u^{s}_{\mathbf{p}} \bar{u}^{s}_{\mathbf{p}}=\slashed{p}+m, &\qquad \qquad& \sum_{s} v^{s}_{\mathbf{p}} \bar{v}^{s}_{\mathbf{p}}=\slashed{p}-m. \\
\end{array}
\end{equation}

The dual field  in terms of $\eta$, is given by 
\begin{equation}\label{dualeta}
\widehat\psi=\eta^{-1}\bar\psi\eta,
\end{equation}
where $\bar{\psi}$ is the Dirac adjoint 
\begin{equation}
\begin{split}
\bar{\psi}&=\psi^{\dagger}\gamma^{0}\\&=\int \frac{d^3\mathbf{p}}{(2\pi)^{3}2\sqrt{m\omega_{\mathbf{p}}}}\sum_s \bigg\{\left(\bar{u}^{s}_{\mathbf{p}}a^{1 s\dagger}_{\mathbf{p}}-\bar{u}^{s}_{\mathbf{p}}\gamma^5a^{2 s\dagger}_{\mathbf{p}}\right)e^{ip\cdot x}+\left(\bar{v}^{s}_{\mathbf{p}}b^{1 s}_{\mathbf{p}}-\bar{v}^{s}_{\mathbf{p}}\gamma^5b^{2 s}_{\mathbf{p}}\right)e^{-ip\cdot x}\bigg\}.
\end{split}
\label{dirac_adjoint}
\end{equation}
Comparing this expression with Eq.(\ref{field_expansion}), the defining action of the operator $\eta$ on the creation and annihilation operators is the change of sign of the operators $a^{2 s}_{\mathbf{p}}$ and $b^{1 s}_{\mathbf{p}}$, summarized as
\begin{equation}
\begin{array}{cccc}
\eta^{-1} a^{j s}_{\mathbf{p}}\eta=(-1)^{j-1} a^{j s}_{\mathbf{p}},\qquad \eta^{-1} b^{j s\dagger}_{\mathbf{p}}\eta= (-1)^{j} b^{j s\dagger}_{\mathbf{p}},
\end{array}
\end{equation}
with $j=1,2$. Transforming the momentum space operators twice with the operator $\eta$ yields
\begin{equation}
\begin{array}{cccc}
\eta^{-1}\eta^{-1} a^{j s}_{\mathbf{p}}\eta\eta=a^{j s}_{\mathbf{p}},\qquad \eta^{-1}\eta^{-1} b^{j s\dagger}_{\mathbf{p}}\eta\eta= b^{j s\dagger}_{\mathbf{p}},
\end{array}
\end{equation}
and therefore, we have $\eta^2=1$ up to an unphysical phase, and consequently $\eta=\eta^{-1}$.
The explicit solution for $\eta$ is 
\begin{equation}
\eta=\exp\left[i\pi\int\frac{d^3\mathbf{p}}{(2\pi)^{3}}\sum_s \left(a^{2 s\dagger}_{\mathbf{p}}a^{2 s}_{\mathbf{p}}+b^{1 s\dagger}_{\mathbf{p}}b^{1 s}_{\mathbf{p}}\right)\right]\equiv e^{i\pi N_{\mathrm{odd}}},
\end{equation}
where we have identified $N_{\mathrm{odd}}$ as the number operator of the particles created by the operators $a^{2 s\dagger}_{\mathbf{p}}$ and $b^{1 s\dagger}_{\mathbf{p}}$. The operator $N_{\mathrm{odd}}$ has integer eigenvalues when applied on particle states, implying that $\eta$ has eigenvalues $\pm 1$, correspondingly.
It can be straightforwardly verified that $\eta$  fulfills the pseudo-Hermiticity of Eq.~(\ref{Lag1}), in light of Eq.~(\ref{pseudo-Hermitian}). 

\section{Canonical quantization}
The equal-time canonical quantization prescription for boson fields is given by
 \begin{equation}\label{can_1}
 \begin{split}
\left[\psi_{\alpha}(\mathbf{x},t),\pi_\psi{}_{\beta}(\mathbf{x}',t)\right]&=\left[\widehat{\psi}_{\alpha}(\mathbf{x},t),\pi_{\widehat{\psi}}{}_{\beta}(\mathbf{x}',t)\right]=i\delta_{\alpha\beta}\delta^{(3)}(\mathbf{x}-\mathbf{x}').
\end{split}
\end{equation} 
with $\alpha,\beta=1,\ldots,4$, and canonical momenta for the field and the dual defined by 
 \begin{equation}
\pi_\psi=\partial_t\widehat\psi=\dot{\widehat{\psi}},\qquad\qquad \pi_{\widehat{\psi}}=\partial_t\psi=\dot{\psi},
\end{equation}
respectively.
The imposition of Eq.(\ref{can_1}) yields the following commutation relations for the momentum-space field operators:
\begin{equation}
\begin{split}
\left[a^{j s}_{\mathbf{p}},a^{k r\dagger}_{\mathbf{p}'}\right]&=(2\pi)^3\delta^{j k}\delta^{r s}\delta^{(3)}(\mathbf{p}-\mathbf{p}'),\qquad
\left[b^{j s}_{\mathbf{p}},b^{k r\dagger}_{\mathbf{p}'}\right]=(2\pi)^3\delta^{j k}\delta^{r s}\delta^{(3)}(\mathbf{p}-\mathbf{p}'),
\end{split}
\label{canonicalrelations}
\end{equation}
with all other commutators vanishing. Notice that the role of the operator $\eta$ is to change the sign of $[a^{2 s}_{\mathbf{p}},a^{2 r\dagger}_{\mathbf{p}'}]$ and $[b^{1 s}_{\mathbf{p}},b^{1 r\dagger}_{\mathbf{p}'}]$ with respect to that which would be obtained using the Dirac adjoint instead.
In addition, the theory features the correct microcausality structure, \emph{i.e.}, the commutator of the field with itself and the commutator of the dual with itself vanish
\begin{equation}
    \comm{\psi_\alpha(x)} {\psi_\beta(y)}=\comm{\widehat{\psi}_\alpha(x)}{ \widehat{\psi}_\beta(y)}=0,
\end{equation}
while the commutator of the field and its dual takes the expected form for a boson field
\begin{equation}\label{micro}
 \begin{split}
&\left[\psi_\alpha(x),\widehat{\psi}_\beta(y)\right]=\delta_{\alpha\beta}\Delta(x-y)=\delta_{\alpha\beta}\int  \frac{d^3\mathbf{p}  }{(2\pi)^{3}2{\omega_{\mathbf{p}}}}\bigg\{ e^{-ip\cdot
   (x- y)}-e^{ip\cdot
   (x- y)}  \bigg\},
    \end{split}
\end{equation}
where $\Delta(x-y)$ is Schwinger's Green function.

The Hamiltonian and momentum operators for our field have the form
\begin{equation}\label{ham}
\begin{split}
H&=:\int d^3\mathbf{x}\bigg(\dot{\widehat{\psi}}\dot{\psi}+\nabla\widehat\psi\cdot\nabla\psi+m^2\widehat\psi\psi\bigg):,\\
\mathbf{P}&=:-\int d^3\mathbf{x}\bigg(\dot{\widehat{\psi}}\nabla\psi+\nabla\widehat{\psi}\dot{ \psi}\bigg):,
\end{split}
\end{equation}
where $:\ :$ stands for bosonic normal ordering. Explicit calculation leads to 
\begin{equation}
\begin{split}
    P^{\mu}&= :\int  \frac{d^3\mathbf{p}}{(2\pi)^{3}}p^{\mu}\sum_{j,s}\left(a^{j s\dagger}_{\mathbf{p}}a^{j s}_{\mathbf{p}}+b^{j s}_{\mathbf{p}}b^{j s\dagger}_{\mathbf{p}}\right):=\int  \frac{d^3\mathbf{p}}{(2\pi)^{3}}p^{\mu}\sum_{j,s}\left(a^{j s\dagger}_{\mathbf{p}}a^{j s}_{\mathbf{p}}+b^{j s\dagger}_{\mathbf{p}}b^{j s}_{\mathbf{p}}\right),
\end{split}
\end{equation}
with $P^{\mu}=(H,\mathbf{P})$. Thus, the energy is bounded from below, making the canonical quantization consistent. Notice that the free normal-ordered Hamiltonian is actually Hermitian when written in terms of the momentum-space operators. 

The Feynman propagator of the free theory  can be straightforwardly calculated 
\begin{equation}
\begin{split}
&[S_F(x-y)]_{\alpha\beta}=\bra{0}T[\psi_\alpha(x)\widehat{\psi}_\beta(y)]\ket{0}\\
&=\bra{0}[\theta(x^0-y^0)\psi_\alpha(x)\widehat{\psi}_\beta(y)+\theta(y^0-x^0)\widehat{\psi}_\beta(y)\psi_\alpha(x)]\ket{0}\\
&=\int\frac{d^4p}{(2\pi)^4}\left[\frac{i\delta_{\alpha\beta}}{p^2-m^2+i\epsilon}\right]e^{-i p\cdot(x-y)},
\end{split}
\end{equation}
and coincides with the propagator expected for a field that satisfies the Klein-Gordon equation. 

\section{Symmetries}
The fields in Eq.(\ref{field_expansion}) describe by construction a spin-1/2 Lorentz covariant theory with boson statistics that satisfies the Klein-Gordon equation. In fact, this statement is indeed very counterintuitive. Thus, in order to show that the SOPHY actually contains particles of spin-1/2, we analyze the rotational properties of the theory.

Since the field belongs to the $(1/2,0)\oplus(0,1/2)$ representation of the RLG, it transforms under rotations as
\begin{equation}\label{rot_1}
    U(R)^{-1}\psi(t,\mathbf{x}) U(R)=e^{-i\boldsymbol{\theta}\cdot \mathbf{J}}\psi(t, R^{-1}\mathbf{x}),
\end{equation}
where the rotation generators have components
\begin{equation}
J^{i}=\frac{1}{2}\epsilon_{ijk}M^{\mu\nu}, \qquad M^{\mu\nu}=\frac{i}{4}[\gamma^\mu,\gamma^\mu].
\end{equation}
In the chiral basis, these generators become
\begin{equation}
\mathbf{J}=\frac{1}{2}\begin{pmatrix}
 \boldsymbol{\sigma} & 0\\
 0 & \boldsymbol{\sigma}
\end{pmatrix}.
\end{equation}
Moreover, if this field describes spin-$1/2$ particles, its momentum space operators must transform under rotations as
\begin{equation}\label{rot_2}
\begin{split}
    U(R)^{-1}a_\mathbf{p}^{js} U(R)&=\sum_{s'}[e^{-i\boldsymbol{\theta}\cdot \boldsymbol{\sigma}/2}]_{ss'}a_{R^{-1}\mathbf{p}}^{js'}\,,\\
        U(R)^{-1}b_\mathbf{p}^{js} U(R)&=\sum_{s'}[e^{-i\boldsymbol{\theta}\cdot \boldsymbol{\sigma}/2}]_{ss'}b_{R^{-1}\mathbf{p}}^{js'}\,.
\end{split}    
\end{equation}
Compatibility between Eqs.(\ref{rot_1},\ref{rot_2}) implies the following conditions \cite{Weinberg:1995mt}:
\begin{gather}
\frac{1}{2}\sum_{s}u^s_{l\boldsymbol{0}}\boldsymbol{\sigma}_{ss'}=\sum_{l'}\mathbf{J}_{ll'}u^{s'}_{l'\boldsymbol{0}}\,, \qquad
-\frac{1}{2}\sum_{s}v^s_{l\boldsymbol{0}}\boldsymbol{\sigma}^*_{ss'}=\sum_{l'}\mathbf{J}_{ll'}v^{s'}_{l'\boldsymbol{0}}\,,\label{rot_3}\\
\frac{1}{2}\sum_{s}\gamma^5 u^s_{l\boldsymbol{0}}\boldsymbol{\sigma}_{ss'}=\sum_{l'}\mathbf{J}_{ll'}[\gamma^5 u^{s'}]_{l'\boldsymbol{0}}\,, \qquad
-\frac{1}{2}\sum_{s}\gamma^5v^s_{l\boldsymbol{0}}\boldsymbol{\sigma}^*_{ss'}=\sum_{l'}\mathbf{J}_{ll'}[\gamma^5v^{s'}]_{l'\boldsymbol{0}}\,.\label{rot_4}
\end{gather}
Using $[\mathbf{J},\gamma^5]=0$, and splitting the spinors into left and right components
\begin{equation}
u^s_{\boldsymbol{0}}=\begin{pmatrix}
 u^s_{L\boldsymbol{0}}\\
 u^s_{R\boldsymbol{0}}
\end{pmatrix},\qquad v^s_{\boldsymbol{0}}=\begin{pmatrix}
 v^s_{L\boldsymbol{0}}\\
 v^s_{R\boldsymbol{0}}
\end{pmatrix},\
\end{equation}
 Eqs.(\ref{rot_3},\ref{rot_4}) become
 \begin{equation}
\sum_{s}(u^s_{L/R\,\boldsymbol{0}})_j\boldsymbol{\sigma}_{ss'}=\sum_{j'}\boldsymbol{\sigma}_{jj'}(u^{s'}_{L/R\,\boldsymbol{0}})_{j'}, \qquad -\sum_{s}(v^s_{L/R\,\boldsymbol{0}})_j\boldsymbol{\sigma}^*_{ss'}=\sum_{j'}\boldsymbol{\sigma}_{jj'}(v^{s'}_{L/R\,\boldsymbol{0}})_{j'}.
 \end{equation}
By considering $(u^s_{L/R\,\boldsymbol{0}})_j$ and $(v^s_{L/R\,\boldsymbol{0}})_j$ as the $(j,s)$ matrix elements of the corresponding matrices $U_{L/R}$, $V_{L/R}$, the above conditions can be written in matrix form as
\begin{equation}\label{rot_5}
U_{L/R}\boldsymbol{\sigma}=\boldsymbol{\sigma} U_{L/R},\qquad -V_{L/R}\boldsymbol{\sigma}^*=\boldsymbol{\sigma} V_{L/R}.
\end{equation}
From Eq.(\ref{u_v}), the explicit form of the $U_{L/R}$ and $V_{L/R}$ matrices is
\begin{equation}
 U_L=U_R=\sqrt{m}\mathbf{1}, V_L=-V_R=-i\sqrt{m}\sigma^2,  
\end{equation}
and therefore Eq.(\ref{rot_5}) is satisfied by virtue of $\sigma^{2}\boldsymbol{\sigma}\sigma^{2}=-\boldsymbol{\sigma}^*$, confirming the rotational covariance of
the SOPHY.

As can be seen explicitly in Eq.~\eqref{Lag1}, the theory possesses a symmetry under global $U(1)$ transformations $\psi\to e^{i\theta}\psi,\ \widehat{\psi}\to e^{-i\theta}\widehat{\psi}$, with $\theta$ being a constant real parameter. The corresponding conserved charge is given by
\begin{equation}\label{charge}
\begin{split}
Q&=:i\int d^3\mathbf{x}\bigg(\widehat{\psi}\dot{\psi}-\dot{\widehat{\psi}}\psi
  \bigg):= :\int  \frac{d^3\mathbf{p}}{(2\pi)^{3}}\sum_{j,s}\left(a^{j s\dagger}_{\mathbf{p}}a^{j s}_{\mathbf{p}}-b^{j s}_{\mathbf{p}}b^{j s\dagger}_{\mathbf{p}}\right):\\
  &= \int  \frac{d^3\mathbf{p}}{(2\pi)^{3}}\sum_{j,s}\left(a^{j s\dagger}_{\mathbf{p}}a^{j s}_{\mathbf{p}}-b^{j s\dagger}_{\mathbf{p}}b^{j s}_{\mathbf{p}}\right),
\end{split}
\end{equation}
which turns out to be explicitly Hermitian, as the normal-ordered Hamiltonian. 

There is actually a larger global symmetry in the free theory; due to the commutation relation between the field and its dual $\comm{\psi_\alpha(x)}{ \widehat{\psi}_\beta(x)} = 0$, obtained from Eq.~\eqref{micro} when $x=y$. The Lagrangian in Eq.~\eqref{Lag1} can be written as 
\begin{equation}
\mathcal{L} = \frac{1}{2} \partial^\mu \Psi^T \Xi\,  \partial_\mu \Psi - \frac{m^2}{2} \Psi^T \Xi \, \Psi,\label{Lag2}
\end{equation}
where we have defined the column matrix
\begin{equation}
\Psi(x) = \begin{pmatrix}
\hat{\psi}^T(x) \\
\psi(x)
\end{pmatrix},
\end{equation}
and $\Xi$ is a $8 \times 8$ matrix, cast in $4 \times 4$ blocks as
\begin{equation}
\Xi = \begin{pmatrix}
0_{4 \times 4} & 1_{4 \times 4} \\
1_{4 \times 4} & 0_{4 \times 4}
\end{pmatrix}=\sigma^1\otimes 1_{4 \times 4}.
\end{equation}
Thus, Eq.~(\ref{Lag2}) is symmetric under global transformations $\Psi \to \Psi' = M \Psi$ with $M^T \Xi M = \Xi$. This group of transformations is isomorphic to $O(8,\mathbb{C})$. Writing $M=e^{\alpha X}$, with $\alpha$ as a complex parameter, the generators $X$ satisfy $X^T\Xi+\Xi X=0$. A suitable basis for the 28 independent generators is 
\begin{equation}
1_{2 \times 2}\otimes A_1,\qquad \sigma^1\otimes A_2,\qquad \sigma^2\otimes A_3,\qquad \sigma^3\otimes S,
\end{equation}
where $A_i, S$, $i,=1,2,3$ are arbitrary antisymmetric and symmetric $4 \times 4$ matrices, respectively.

The discrete transformations of the field and its dual under parity ($P$), charge conjugation ($C$), and time reversal ($T$) follow from the transformation properties of their momentum-space operators
\begin{gather}
\mathrm{P}a^{js\dagger}_{\mathbf{p}}\mathrm{P}^{-1}=-i(-1)^{j-1} a^{js\dagger}_{-\mathbf{p}},\quad\mathrm{P}b^{js\dagger}_{\mathbf{p}}\mathrm{P}^{-1}=-i(-1)^{j-1}b^{js\dagger}_{-\mathbf{p}},\nonumber\\
\mathrm{C}a^{js\dagger}_{\mathbf{p}}\mathrm{C}^{-1}= -b^{js\dagger}_{\mathbf{p}},\quad\quad\quad\mathrm{C}b^{js\dagger}_{\mathbf{p}}\mathrm{C}^{-1}=a^{js\dagger}_{\mathbf{p}},\\
\mathrm{T}a^{js\dagger}_{\mathbf{p}}\mathrm{T}^{-1}=2s a^{j(-s)\dagger}_{-\mathbf{p}},\quad\quad\quad\mathrm{T}b^{js\dagger}_{\mathbf{p}}\mathrm{T}^{-1}=  2s b^{j(-s)\dagger}_{-\mathbf{p}},\nonumber
\end{gather} 
and are given by
\begin{gather}
\mathrm{P}\psi(x)\mathrm{P}^{-1}=i\gamma^0\psi(\mathcal{P}x),\quad \mathrm{P}\widehat{\psi}(x)\mathrm{P}^{-1}=-i\widehat{\psi}(\mathcal{P}x)\gamma^0,\nonumber\\
\mathrm{C}\psi(x)\mathrm{C}^{-1}=  \mathcal{C} \widehat{\psi}^{\,T},\qquad \mathrm{C}\widehat{\psi}\mathrm{C}^{-1}=-\psi^T\mathcal{C},\\
\mathrm{T}\psi(x)\mathrm{T}^{-1}=\mathcal{C}\gamma^5 \psi(\mathcal{T}x),\qquad \mathrm{T}\widehat{\psi}\mathrm{T}^{-1}=-\widehat{\psi}(\mathcal{T}x) \gamma^5\mathcal{C},\nonumber
\end{gather} 
with $\mathcal{P}=\mathrm{diag}(1,-1,-1,-1)$, $\mathcal{T}=\mathrm{diag}(-1,1,1,1)$, and $\mathcal{C}=-i \gamma^2\gamma^0$ in the chiral representation.
The corresponding pseudo-Hermitian bilinears transform under $P$ as
\begin{equation}
\begin{split}
\mathrm{P}\widehat{\psi}\psi\mathrm{P}^{-1}&=\widehat{\psi}\psi,\nonumber\\
\mathrm{P}\widehat{\psi}i\gamma^5\psi\mathrm{P}^{-1}&=-\widehat{\psi}i\gamma^5\psi,\nonumber\\
\mathrm{P}\widehat{\psi}\gamma^\mu\psi\mathrm{P}^{-1}&=\mathcal{P}^\mu{}_\nu\widehat{\psi}\gamma^\nu\psi,\nonumber\\
\mathrm{P}\widehat{\psi}\gamma^\mu\gamma^5\psi\mathrm{P}^{-1}&=-\mathcal{P}^\mu{}_\nu\widehat{\psi}\gamma^\nu\gamma^5\psi,\nonumber\\
\mathrm{P}\widehat{\psi}\sigma^{\mu\nu}\psi\mathrm{P}^{-1}&=\mathcal{P}^\mu{}_\lambda\mathcal{P}^\nu{}_\rho\widehat{\psi}\sigma^{\lambda\rho}\psi,
\end{split}
\end{equation}
where $\sigma^{\mu\nu}\equiv2M^{\mu\nu}=i[\gamma^\mu,\gamma^\nu]/2$. Similarly, the transformations of the bilinears under $C$ and $T$ are
\begin{equation}
\begin{split}
\mathrm{C}\widehat{\psi}\psi\mathrm{C}^{-1}&=\widehat{\psi}\psi,\nonumber\\
\mathrm{C}\widehat{\psi}i\gamma^5\psi\mathrm{C}^{-1}&=\widehat{\psi}i\gamma^5\psi,\nonumber\\
\mathrm{C}\widehat{\psi}\gamma^\mu\psi\mathrm{C}^{-1}&=-\widehat{\psi}\gamma^\mu\psi,\nonumber\\
\mathrm{C}\widehat{\psi}\gamma^\mu\gamma^5\psi\mathrm{C}^{-1}&=+\widehat{\psi}\gamma^\mu\gamma^5\psi,\nonumber\\
\mathrm{C}\widehat{\psi}\sigma^{\mu\nu}\psi\mathrm{C}^{-1}&=-\widehat{\psi}\sigma^{\lambda\rho}\psi,
\end{split}
\end{equation}
and
\begin{equation}
\begin{split}
\mathrm{T}\widehat{\psi}\psi\mathrm{T}^{-1}&=\widehat{\psi}\psi,\nonumber\\
\mathrm{T}\widehat{\psi}i\gamma^5\psi\mathrm{T}^{-1}&=-\widehat{\psi}i\gamma^5\psi,\nonumber\\
\mathrm{T}\widehat{\psi}\gamma^\mu\psi\mathrm{T}^{-1}&=-\mathcal{T}^\mu{}_\nu\widehat{\psi}\gamma^\nu\psi,\nonumber\\
\mathrm{T}\widehat{\psi}\gamma^\mu\gamma^5\psi\mathrm{T}^{-1}&=-\mathcal{T}^\mu{}_\nu\widehat{\psi}\gamma^\nu\gamma^5\psi,\nonumber\\
\mathrm{T}\widehat{\psi}\sigma^{\mu\nu}\psi\mathrm{T}^{-1}&=-\mathcal{T}^\mu{}_\lambda\mathcal{T}^\nu{}_\rho\widehat{\psi}\sigma^{\lambda\rho}\psi.
\end{split}
\end{equation}
Therefore, the free theory is invariant under each discrete symmetry simultaneously, and also under $CPT$.

Regarding interactions, since our fields have mass dimension one, it is possible to introduce renormalizable quartic self-interactions, in contrast to Dirac theory. The simplest  $C$, $P$ and $T$ invariant pseudo-Hermitian interactions are given by
\begin{equation}
\begin{split}
\mathcal{L}_{\text{int}}=&   \frac{\lambda_{1}}{2}\left(\widehat{\psi}\psi\right)^{2}
+ \frac{\lambda_{2}}{2}\left( \widehat{\psi}\gamma^{5}\psi\right)  \left(\widehat{\psi}\gamma^{5}\psi\right)+ \frac{\lambda_{3}}{2}\left( \widehat{\psi}M^{\mu\nu}\psi\right)  \left(\widehat{\psi}M_{\mu\nu}\psi\right).
\label{Lag3}
\end{split}
\end{equation}
Other self-interactions can be written in terms of the above three through a Fierz transformation. These quartic interactions are
analogous to those studied for fermion fields in \cite{Vaquera-Araujo:2012jlk,Vaquera-Araujo:2013bwa} in the context of the naive Hermitian second-order theory for fermions, where it was shown that this class of theories has a rich set of renormalization group equations, compared to their scalar counterparts. Thus, spin-$1/2$ SOPHIES constitute  valuable theoretical playgrounds for renormalization studies.  Besides, due to its dimensionality, this new field can not couple to the quarks and leptons of the Standard Model in a renormalizable way. If the field $\psi$ is not charged under the Standard Model gauge symmetries, it can be identified as a WIMP dark matter candidate. The natural interaction of this field with the SM particles is through a Higgs portal of the form
\begin{equation}
\mathcal{L}_{\psi H}=   \lambda_{H}\left(\widehat{\psi}\psi\right)H^\dagger H\label{LagHiggs},
\end{equation}
where $H$ is the SM Higgs doublet.
In this case the dark matter phenomenology is determined by the mass of the field $m$ and its coupling $\lambda_{H}$ in a similar way as the complex scalar WIMP DM ~\cite{Feng:2014vea,Arcadi:2017kky}, since the field satisfies the Klein-Gordon equation. Another possible DM model is through the effective interaction
\begin{equation}
\mathcal{L}_{\psi h}=   ig_{h}\left(\widehat{\psi}\gamma^\mu\partial_\mu\psi\right)h\label{LagEff},
\end{equation}
with $h$ as the physical Higgs scalar. This interaction has been recently studied in the context of a spin-$1/2$ fermion model in \cite{deGracia:2024umr}.

\section{Conclusions}

Summarizing, in this work, we have presented the canonical quantization of a massive boson field transforming under the $(\tfrac{1}{2},0)\oplus(0,\tfrac{1}{2})$ representation of the RLG and obeying the Klein-Gordon equation. The quantization of the free theory is consistent, and the usual spin-statistics connection is absent due to the redefinition of the field dual. The resulting formalism is surprisingly simple, as it is directly built with two independent Dirac fields, and it belongs to a class of second order pseudo-Hermitian theories, SOPHIES for short, that include those studied in \cite{LeClair:2007iy,Ahluwalia:2023slc,Ferro-Hernandez:2023ymz}, where Hermiticity is replaced by the less stringent relation displayed in Eq.(\ref{pseudo-Hermitian}). 

\vspace{6pt} 






\funding{This work was supported  by the Secretar\'ia de Ciencia, Humanidades, tecnolog\'ia e Innovacion  (SECIHTI) through the project IxM 749. A.d.l.C.R.-P. acknowledges SECIHTI national scholarships. I. D-S. was supported by the SECIHITI program ``Estancias Posdoctorales por M\' exico''.}



\dataavailability{Data sets generated during the current study are available from the corresponding author on reasonable request.}

\abbreviations{Abbreviations}{
The following abbreviations are used in this manuscript:
\\

\noindent 
\begin{tabular}{@{}ll}
SOPHY & Second-Order Pseudo-Hermitian theory\\
RLG & Restricted
Lorentz Group\\
WIMP & Weakly Interacting Massive Particle
\end{tabular}
}


\reftitle{References}

\end{document}